\documentclass[12pt]{article}
\newcommand{\Comment}[1]{{}}
\usepackage[textwidth = 450 pt, textheight = 630 pt]{geometry}
\usepackage{amssymb,euscript,amsmath,amsfonts}
\usepackage{tikz,tikz-cd,url}
\usetikzlibrary{decorations.pathreplacing,calligraphy}
\usepackage{xcolor}
\usepackage{graphicx}
\usepackage{color}
\usepackage{bbm}
\usepackage{tensor}
\usepackage{slashed}

\definecolor{MyDarkBlue}{rgb}{0.15,0.15,0.45}
\usepackage[linktocpage=true]{hyperref}
\hypersetup{
colorlinks=true,
citecolor=MyDarkBlue,
linkcolor=MyDarkBlue,
urlcolor=MyDarkBlue,
pdfauthor={Neil Lambert },
pdftitle={ },
pdfsubject={hep-th}
}

\usepackage{hypernat}
\usepackage{physics}
\usepackage{accents}
\usepackage{bm}
\newcommand{\inv}[1]{\frac{1}{#1}}
\newcommand{\mn}{\mu\nu}
\newcommand{\fint}{\int dt d^4x\,}
\DeclareMathOperator{\eff}{eff}

\renewcommand{\frak}[1]{\mathfrak{#1}}
\newcommand{\brho}{\rho^{\dag}}
\renewcommand{\cal}[1]{\mathcal{#1}}
\newcommand{\bchi}{\chi^{\dag}}
\newcommand{\vphi}{X}
\newcommand{\balpha}{\Bar{\alpha}}
\newcommand{\bbeta}{\Bar{\beta}}
\newcommand{\bvphi}{{X}^\dag}
\newcommand{\brac}[1]{\left(#1\right)}

\begin{document}

\renewcommand{\thefootnote}{\fnsymbol{footnote}}

   \vspace{1.8truecm}

 \centerline{\LARGE \bf {\sc RG Flows and Symmetry Enhancement in }} \vskip 12pt
 \centerline{\LARGE \bf{\sc  Five-Dimensional Lifshitz Gauge Theories}}

\centerline{\LARGE \bf {\sc  }} \vspace{2truecm} \thispagestyle{empty} \centerline{
    {\large {{\sc Neil~Lambert}}}\footnote{E-mail address: \href{mailto:neil.lambert@kcl.ac.uk}{\tt neil.lambert@kcl.ac.uk} } and  {\large {{\sc Joseph~Smith}}}\footnote{E-mail address: \href{mailto:joseph.m.smith@kcl.ac.uk}{\tt joseph.m.smith@kcl.ac.uk} }
  }

\vspace{1cm}
\centerline{{\it Department of Mathematics}}
\centerline{{\it King's College London }} 
\centerline{{\it The Strand }} 
\centerline{{\it  WC2R 2LS, UK}} 

\vspace{1.0truecm}

\thispagestyle{empty}

\centerline{\sc Abstract}
\vspace{0.4truecm}
\begin{center}
\begin{minipage}[c]{360pt}{
    \noindent}
    
 Lagrangian gauge theories with a $z=2$ Lifshitz scaling provide a family of interacting, asymptotically free five-dimensional field theories. We examine a broad class of these theories, including some of their quantum properties, extending previous results to include matter. We present   no-go theorems that, in the absence of constraints, the class of theories we consider cannot admit a spinorial supersymmetry or Galilean boost symmetry. However, we argue that there exist renormalization group flows whose fixed points can admit supersymmetry and boosts, {\it i.e.} super-Schr\"odinger symmetry. We also present examples of Lifshitz gauge theories with a scalar supersymmetry. 

\end{minipage}
\end{center}

\newpage 
 \section{Introduction}\label{sect: Intro}
 
Non-Lorentzian Lifshitz gauge theories provide an interesting class of non-trivial interacting gauge theories in five dimensions which are perturbatively renormalizable and asymptotically free. Such theories have a quadratic kinetic term, so as to avoid ghosts, but higher derivative spatial gradients. This breaks Lorentz invariance but leads to their desirable high energy behaviour. They have been studied from a variety of perspectives,  often with the view to recovering Lorentzian theories in the infrared. For a sample of this literature see \cite{Horava:2008jf,Iengo:2009ix,Iengo:2010xg,Kanazawa:2014fla}.

Here we wish to revisit these theories with a different goal in mind. It is well known that null reduction of a six-dimensional relativistic field theory leads to a five-dimensional Galilean theory, {\it i.e.} a non-relativistic theory with rotations and boosts, as well as a further central $U(1)$ symmetry arising from the compact null momentum (whose conserved charge is often identified with particle number or mass). If  we instead start with a conformal field theory then we find  Schr\"odinger theories which include a $z=2$ Lifshitz scaling symmetry as well as a special conformal transformation. Null reduction of superconformal field theories  lead to super-Schr\"odinger theories (again with a central $U(1)$ extension).

Indeed, five-dimensional
 super-Schr\"odinger Lagrangian gauge theories have been explicitly  constructed in \cite{Lambert:2018lgt,Lambert:2020jjm,Lambert:2019fne} as null reductions of six-dimensional Lorentzian conformal field theories.
 These theories have the unusual feature that they contain Lagrange multiplier fields which imply that the physical dynamics takes place on the moduli space of (anti-)self-dual gauge fields on ${\mathbb R}^4$, and like all five-dimensional gauge theories, they admit a topological $U(1)$ symmetry whose charge is the instanton number. They therefore provide a Lagrangian field-theoretic realization of the DLCQ descriptions of six-dimensional superconformal field theories \cite{Aharony:1997th,Aharony:1997an}. Furthermore, $\Omega$-deformed versions of these theories enjoy a novel $SU(1,3)$ spacetime symmetry and can reproduce aspects of uncompactified six-dimensional superconformal field theories \cite{Lambert:2019jwi,Lambert:2020zdc}.

 Thus it is natural to ask if we can find unconstrained supersymmetric  Lagrangian gauge theories with a $z=2$ Lifshitz scale symmetry and boosts that have admit a weakly coupled Lagrangian description as a deformation  of free fields.  We will argue that the answer to this question is no: there are no supersymmetric  or Galilean boost invariant Lifshitz gauge theories unless one allows for Lagrange multiplier fields that heavily constrain the system and exclude a weakly coupled limit. 
 
 However, we will argue that one can start from a non-supersymmetric Lifshitz gauge theory in the UV and include relevant deformations that cause the theory to flow in the IR to super-Schr\"odinger theories, {\it i.e.} supersymmetry and boosts can arise as enhanced symmetries in the IR. In particular we will recover precisely the Lagrangian field theories mentioned above involving constraints. 
Thus our interest here is not to employ five-dimensional Lifshitz gauge theories as UV descriptions of five-dimensional Lorentzian theories, but rather as   perturbative UV descriptions of six-dimensional superconformal theories!

Lastly, we note that when we referred to supersymmetry above we meant Fermionic symmetries that transform as a spinor under the four-dimensional rotation group and which close on the Hamiltonian. The Fermions in such theories are then also spinors of the rotation group. However, without Lorentz invariance there is no spin-statistics theorem and one can contemplate scalar or vector Fermions. We will explicitly see that one can  construct supersymmetric Lifshitz gauge theories where the supersymmetry is a scalar and the Fermions vectors of the four-dimensional rotation group. These actions are similar to the Parisi-Sourlas construction \cite{Parisi:1982ud}. Indeed, other non-gauge theory scalar supersymmetric theories have been constructed and studied in \cite{Chapman:2015wha,Arav:2019tqm}.
 
 The rest of this paper is organised as follows. In section two we will review the class of five-dimensional Lifshitz gauge theories that we wish to consider. In section three we compute the one-loop $\beta$-function of the coupling $g$  and find agreement with previous results \cite{Horava:2008jf} which were obtained in the case without matter. However, we also consider the one-loop contributions of scalar and Fermionic matter fields to the $\beta$-function. In section four we argue that we there do not exist supersymmetric (with spinorial generators) Lifshitz gauge theories, or Lifshitz gauge theories  with Galilean boosts, without including Lagrange multiplier fields. The problem arises as a conflict between the supersymmetry algebra and the dispersion relation and is apparent already in the free theory. In section five we argue that by adding relevant terms to a Lifshitz gauge theory one can induce RG flows that lead to supersymmetric theories and theories with boosts. In section six we give a construction of supersymmetric Lifshitz gauge theories coupled to matter but where the supercharge is a scalar and the Fermions vectors under the four-dimensional rotation group. In section seven we give our conclusions. In addition details of several computations have been given in the appendix.

\section{ Lifshitz Gauge Theories}
 \label{sect: Lifshitz}
 
Let us begin by considering a Bosonic Lifshitz gauge theory in flat space of the form 
\begin{align} \label{eq: action 1}
 S & = \inv{2g^2} \fint\Tr  \brac{ F_{t\mu} F_{t\mu} - \lambda^2 D_\nu F_{\nu\mu}D_\rho F_{\rho\mu} } \ .
 \end{align}
 where $\mu,\nu =1,2,3,4$ and repeated indices are summed over. 
 As well as the obvious translation and $SO(4)$ rotational symmetries, this action is invariant under the $z=2$ Lifshitz scaling
 \begin{align} \nonumber
     A'_{\mu}(t,x) &= s A_{\mu}(s^2 t, s x) \\
     A'_t(t,x) &= s^2 A_t(s^2 t, s x) \ .
\end{align}
 As the scaling of time and space under the transformation differ, it is natural to fix our dimensions such that
 \begin{equation}
     [E] = [T]^{-1} = [L]^{-2} \ ,
 \end{equation}
 where $E,T,L$ refer to energy, time and distance respectively.
 We will take $[L]^{-1}$ to define the Lifshitz mass dimension of our theory ({\it i.e.} we take $[L]^{-1} = 1$). The scaling dimensions of our fields then match the (Lifshitz) mass dimension, as occurs in relativistic field theories with scaling symmetry. Any action we write down that has no dimensionful parameters will then automatically be invariant under Lifshitz scalings.

We could ask if there are any other terms we can add to the action that are consistent with the scaling symmetry. We will only consider terms that are parity-even. An obvious candidate is $F_{t\mu} D_{\nu} F_{\nu\mu}$. A brief calculation using the time-space-space component of the Bianchi identity gives
\begin{equation}
    {\rm Tr}\brac{F_{t\mu} D_{\nu} F_{\nu\mu}} = \partial_{\nu} \brac{\Tr \brac{F_{t\mu} F_{\nu\mu} }} - \frac1{4} \partial_t \brac{\Tr \brac{F_{\mn} F_{\mn}}} \ ,
\end{equation}
so we see this term is non-dynamical. We can also consider different contractions of indices in the $(DF)^2$ term. It will be helpful to first consider this question in the Abelian theory. The spatial Bianchi identity can then be used to show that the only non-vanishing parity-even terms of the form (here $\beta$ is an $SO(4)$-invariant tensor)
\begin{equation}
  S_{dF Abelian}=  \fint \beta_{\mu_1 ... \mu_6} \partial_{\mu_1} F_{\mu_2 \mu_3} \partial_{\mu_4} F_{\mu_5 \mu_6}\
\end{equation}
 are proportional to, neglecting total derivative terms,
\begin{equation}
 S_{dF Abelian}   \propto \fint \partial_{\nu} F_{\nu\mu} \partial_{\rho} F_{\rho\mu} \ .
\end{equation}
In going to the non-Abelian theory, the only thing that can change is the possibility of commutator terms of the form
\begin{equation}
  S_{F^3}=     -i\fint \kappa_{\mu_1 ... \mu_6} \Tr (F_{\mu_1 \mu_2} [F_{\mu_3 \mu_4}, F_{\mu_5 \mu_6}]) \ .
\end{equation}
Imposing our parity-even constraint, the only non-vanishing term is
\begin{equation}
 S_{F^3}=  -\frac{i}{2}\kappa \fint\Tr ( F_{\mn}[ F_{\nu \rho}, F_{\rho\mu}] ) \ ,
\end{equation}
for some dimensionless parameter $\kappa$. Thus the general parity-even extension of (\ref{eq: action 1}) is \begin{equation}
    \Tilde{S} = \inv{2g^2}  \fint \Tr\brac{ F_{t\mu} F_{t\mu} - \lambda_1^2 D_{\nu} F_{\nu\mu} D_{\rho} F_{\rho\mu} - \frac12\lambda_2^2 D_{\mu} F_{\nu \rho} D_{\mu} F_{\nu \rho}-\frac{i\kappa}{2}   F_{\mn}[ F_{\nu \rho}, F_{\rho\mu}]  } \ .
\end{equation}
Next we observe that
\begin{align}
    -i \Tr\brac{F_{\mn} [F_{\nu\rho}, F_{\rho\mu}]} =&   \Tr \brac{D_{\nu} F_{\nu\mu} D_{\rho} F_{\rho\mu}}-\inv{2} \Tr\brac{D_{\mu}F_{\nu \rho} D_{\mu}F_{\nu \rho}} \nonumber\\
     & - \partial_{\nu} \brac{\Tr \brac{F_{\mu \rho} D_{\mu} F_{\rho\nu} - F_{\mn} D_{\rho} F_{\rho\mu}}} \ .
\end{align}
Therefore, up to a boundary term, the three coefficients $\lambda_1,\lambda_2$ and $\kappa$ are not independent. Rather the action is invariant  under the shift
\begin{align}
\lambda_1^2 \to \lambda_1^2 -g^2\alpha\qquad \lambda^2_2\to \lambda_2^2+ g^2\alpha \qquad \kappa \to\kappa +\alpha\ .
\end{align}
We  find it useful to use  to  remove the third term, {\it i.e.} set $\lambda_2=0$, so as to have a unique gradient term. In this case we simply denote $\lambda=\lambda_1$.
In such a parameterization, in order to keep the energy bounded from below,    we need to   restrict  $\kappa$ to the range   $0\leq \kappa \leq \lambda^2/g^2$.  It is an important question as to whether or not  $\kappa$ can be so restricted at the quantum level. Unfortunately, answering this is beyond the scope of this paper.

If we want to make contact with supersymmetric theories, we will have to couple the gauge theory to matter fields whose actions are invariant under Lifshitz scaling. The fields are also required to be $\frak{g}$-valued. The simplest field that can be considered is a complex scalar field $X$; we will take its action to be
\begin{equation}
    S_X = \inv{g^2}  \fint \Tr \brac{D_t X^{\dag} D_t X - \lambda_X^2 D^2 X^{\dag} D^2 X} \ .
\end{equation}
It is possible to augment this action with a sixth-order scalar potential and retain scale-invariance, but we won't consider the effect of this here.

As our theory is not Lorentz-invariant, the usual rigidity enforced by the spin-statistics theorem does not apply. This allows us to consider an anticommuting complex scalar $\psi$, also $\frak{g}$-valued, with action
\begin{equation}
    S_{\psi} = \inv{g^2}  \fint \Tr \brac{\psi^{\dag} D_t \psi - \lambda_{\psi}^2 D_{\mu} \psi^{\dag} D_{\mu} \psi} \ .
\end{equation}
We will also work with more familiar Fermions in a spinor representation of $SO(4)$. We will take these to have the same free action as the anticommuting scalar, with conjugation acting on both gauge and spinor indices. Note that the spinorial properties of $\psi$ play essentially no role in the action. We will denote both types of Fermion by $\psi$, since we will never use both simultaneously and they largely play the same role. We will also consider scalars Bosons and Fermions in other representations of the gauge group.

It is important to note that without Lorentz symmetry each field can have its own $\lambda$. If all the $\lambda$'s are equal then one can rescale time to set $\lambda=1$, but in general there is no reason to expect this. 

\subsection{Solutions of the Field Equations}
\label{sect: field equations}

Let us return to the pure gauge theory. 
Varying $A_{\mu}$ and $A_t$ in (\ref{eq: action 1}) give the field equations
\begin{gather} \nonumber 
    D_t F_{t\mu} + \lambda^2 D^2 D_{\nu} F_{\nu\mu} - 2i\lambda^2 [F_{\mn}, D_{\rho} F_{\rho\nu}] = 0 \\ \label{eq: non-abelian field equations}
    D_{\mu} F_{t\mu} = 0 \ .
\end{gather}
First, let's consider the free theory. Rescaling the gauge field to \footnote{Here we use $M$ to denote both temporal and spatial indices.}
\begin{equation}
    A_M(t,x) = g \,a_M(t,x)
\end{equation}
and taking $g\to 0$ gives the equations
\begin{gather} \nonumber 
    \partial_t^2 a_{\mu} + \lambda^2 \partial^4 a_{\mu} - \partial_{\mu} \partial_t a_t - \lambda^2 \partial_{\mu} \partial^2 \partial_{\nu} a_{\nu} = 0 \\ \label{eq: abelian field equations}
    \partial_t \partial_{\mu} a_{\mu} - \partial^2 a_t = 0\ .
\end{gather}
In this limit our gauge transformations become Abelian,
\begin{equation}
    a_M \to a_M' = a_M + \partial_M \alpha \ .
\end{equation}
We can use this gauge freedom to fix $a_t = 0$. This still leaves us with the potential to make further time-independent transformations. These can be used to set
\begin{equation}
    \partial_{\mu} a_{\mu} (t=0,x) = 0\ ,
\end{equation}
on the surface $t=0$. However, when $a_t = 0$ the second equation in (\ref{eq: abelian field equations}) tells us that $\partial_{\mu} a_{\mu}$ is $t$-independent, so fixing it at $t=0$ causes it to vanish for all $t$. The remaining equation in (\ref{eq: abelian field equations}) is then
\begin{equation}
    \partial_t^2 a_{\mu} + \lambda^2 \partial^4 a_{\mu} = 0 \ .
\end{equation}
We can decompose this into the Fourier modes
\begin{equation}
    a_{\mu} = \zeta_{\mu} e^{-i(E t - k_{\nu} x_{\nu})}\ ,
\end{equation}
where
\begin{equation}
    k_{\mu} \zeta_{\mu} = 0\ ,
\end{equation}
and
\begin{equation}
    E^2 = \lambda^2 k^4 \ .
\end{equation}
A similar analysis for the matter fields $X$ and $\chi$ leads to the same dispersion relation but with different parameters $\lambda_X$ and $\lambda_\chi$. Thus perturbatively we find a unique ground state  admitting plane-wave excitations which can then be quantized following standard methods.

As with four-dimensional gauge theories, the interacting theory has a much richer structure than the free theory. Though we can't solve (\ref{eq: non-abelian field equations}) directly, we can make headway by considering static field configurations for which
\begin{equation}
    F_{t\mu} = 0 \ .
\end{equation}
The field equations then reduce to
\begin{equation}
    D^2 D_{\nu} F_{\nu\mu} - 2i[F_{\mn}, D_{\rho} F_{\rho\nu}] = 0 \ ,
\end{equation}
which is still a challenge to solve in general. However, we see that the equation is satisfied whenever the gauge field's spatial components obey the reduced equation
\begin{equation}\label{4deom}
    D_{\mu} F_{\mn} = 0 \ ,
\end{equation}
which means any solution of the field equations of four-dimensional Euclidean Yang-Mills automatically solve the full field equations if they satisfy the static condition. If we work in the $A_t = 0$ gauge, this condition becomes simply 
\begin{equation}
    \partial_t A_{\mu} = 0 \ .
\end{equation}
Since solutions of the four-dimensional theory can be taken to not depend on $t$, we can always ensure the condition is met.

The Hamiltonian of the theory is
\begin{align}
    H & = \frac{1}{2g^2} \int d^4x \, \Tr\brac{ g^2 \Pi_{\mu} \Pi_{\mu} + \lambda^2 D_\nu F_{\nu\mu}D_\rho F_{\rho\mu }+ 2g^2 A_tD_\mu \Pi_\mu } 	\ ,
\end{align}
where
\begin{equation}
    \Pi_\mu  = \inv{g^2}F_{t\mu} \ .
\end{equation}
Working in the gauge $A_t = 0$ and imposing $D_{\mu} \Pi_{\mu}$ as a constraint on our field configurations, we see that $H \geq 0$, with equality only if both $\Pi_{\mu}$ and $D_{\nu}F_{\nu\mu}$ vanish. This means that any solution of the form (\ref{4deom}) is a ground state of the theory.

\subsection{Relevant Deformation}

We can deform our theory with the relevant operator
\begin{equation}
    \Delta S_1 = - \frac{M^2}{2g^2}  \fint \Tr\brac{F_{\mn} F_{\mn}} \ ,
\end{equation}
for which the deformed Hamiltonian, after imposing the Gauss law constraint, is
\begin{equation}
    H = \inv{2g^2} \int d^4x  \Tr \brac{ g^2 \Pi_{\mu} \Pi_{\mu} + \lambda^2 D_{\nu} F_{\nu\mu} D_{\rho} F_{\rho\mu} + M^2 F_{\mn} F_{\mn}} \ .
\end{equation}
As before, the lowest energy states necessarily have
\begin{equation}
    \Pi_{\mu} = D_{\nu} F_{\nu\mu} = 0 \ .
\end{equation}
If the field configuration lies in the $k$-instanton sector, then the Bogomoln'yi bound implies that its energy satisfies
\begin{equation}
    E \geq \frac{8\pi^2 M^2 |k|}{g^2}\ ,
\end{equation}  where $k$ is the spatial instanton number:
\begin{align}
k = \frac{1}{32\pi^2} \int d^4x \, \epsilon_{\mn\rho\sigma}  \Tr\brac{F_{\mn} F_{\rho\sigma}} \ .
\end{align}
The bound is only saturated if
\begin{equation}
    F_{\mn} = \begin{cases}
    \inv{2} \epsilon_{\mn\rho\sigma} F_{\rho\sigma}, & k>0 \\
    -\inv{2} \epsilon_{\mn\rho\sigma} F_{\rho\sigma}, & k<0
    \end{cases} \ ,
\end{equation}
{\it i.e.} the field is in an (anti-)instantonic configuration. Next, suppose we add a further deformation that only depends on the topological properties of our fields,
\begin{equation}
    \Delta S_2 =  \frac{\theta M^2}{4g^2}  \fint \epsilon_{\mn\rho\sigma} \Tr\brac{F_{\mn} F_{\rho\sigma}} \ .
\end{equation}
The change in the Hamiltonian can be directly evaluated, giving
\begin{align} \nonumber
    \Delta H &= -\frac{\theta M^2}{4g^2}  \int d^4x \, \epsilon_{\mn\rho\sigma} \Tr\brac{F_{\mn} F_{\rho\sigma}} \\
    &= -\frac{8\pi^2 \theta M^2 k}{g^2}\ ,
\end{align}
which only depends on the topological sector in which the solution is found. We can choose the coupling of this term to be $\theta = 1$, so our bound becomes
\begin{equation}
    E \geq \frac{8\pi^2 M^2}{g^2} ( |k| - k) \ .
\end{equation}
We see that the energy vanishes for the both the vacuum and instantons, but is positive and proportional to $M^2$ for all other static solutions. Taking the limit $M\to \infty$ then decouples the massive solutions, leaving only the vacuum and instantons as ground states. More generally, the only finite-energy solutions of the theory are instantonic profiles with time-dependent moduli,
so the dynamics of the theory reduce to a non-linear sigma model on instanton moduli space.

\section{One-Loop Renormalization Properties}

Let us compute the effects of perturbations about a background field $A_M\to A_M+ga_m$ where we treat $a_M$ as the quantum field to be integrated over in the path-integral.\footnote{In this section we will work in Euclidean signature to simplify the calculations.} The background field's 1PI effective action, $\Gamma[A]$, is \cite{Abbott:1981ke}
\begin{equation} \label{eq: 1PI}
    e^{-\Gamma[A]} = \int Da\, e^{-S[A + g a]}\ ,
\end{equation}
 with $A_M$ chosen such that $\Gamma$ is extremised. At the 1-loop level it is self-consistent to relax this slightly and require that $A_M$ minimises the classical action, as the difference will only contribute at higher loops. As it stands (\ref{eq: 1PI}) is ill-defined due to the gauge freedom of $a$; we need to specify a gauge fixing procedure to actually compute the path integral. Our route forward will mirror the gauge fixing used in \cite{Casarin:2017xez}. Let $G[a]$ be a gauge-fixing condition, $H$ be an operator that may depend on $A_M$, and $\alpha$ the parameter for a gauge transformation of $a_M$. The gauge fixed path integral is then
 \begin{align} \nonumber
     e^{-\Gamma[A]} &= \int Da D\omega \, \det\brac{\frac{\delta G}{\delta \alpha}} \sqrt{\det H} \delta[G - \omega] \exp\brac{-S[A+ g a] - \inv{2}  \fint \Tr\brac{ \omega H \omega} } \\ 
     &= \int Da\, \det\brac{\frac{\delta G}{\delta \alpha}} \sqrt{\det H} \exp\brac{-S[A + g a] - \inv{2} \fint \Tr \brac{ G H G}} \ .
 \end{align}
 The natural choice of gauge fixing condition is Coulomb gauge,
 \begin{equation}
     G[a] = D_{\mu} a_{\mu} \ ,
 \end{equation}
 and we take  $H$ to be
 \begin{equation}
     H = - \lambda^2 \xi D_\mu D_\mu
 \end{equation}
 for some gauge fixing parameter $\xi$.
 However, it will prove incredibly convenient to choose $\xi=1$ to simplify the propagator and hence we will do that in what follows. Note that in this section all gauge-covariant derivatives are taken with respect to the background field $A$. 
 As
 \begin{equation}
     \delta G = D_{\mu} (D_{\mu} \alpha - i g [a_{\mu}, \alpha])\ ,
 \end{equation}
the Fadeev-Popov determinant at 1-loop can be approximated as
\begin{equation}
    \det \brac{\frac{\delta G}{\delta \alpha}} \approx \det\brac{D^2} \ .
\end{equation}
Expanding the action in powers of $g$ gives
\begin{equation}
    S[A + g a] + \inv{2} \fint \Tr\brac{G H G} = S[A] + \inv{2} \fint \Tr \brac{ a_M \Delta_{MN} a_N} + O(g)\ ,
\end{equation}
for some $A_M$-dependent operator $\Delta$. Neglecting field-independent constants, the 1-loop 1PI effective action is then
\begin{equation} \label{eq: 1PI 2}
    \Gamma[A] = S[A] + \inv{2} \Tr \ln \Delta - \frac{3}{2} \Tr \ln\brac{-D^2}\ .
\end{equation}
A short computation gives
\begin{gather}  \nonumber
    \Delta_{tt} = - D^2 \\ \nonumber
    \Delta_{t\mu} = D_{\mu} D_t + i E_{\mu}^{adj.} \\ \nonumber
    \Delta_{\mu t} = D_t D_{\mu} - i E_{\mu}^{adj.} \\ \nonumber
    \Delta_{\mn} = - \delta_{\mn} D_t^2 + \lambda^2 \big( \delta_{\mn} D^4 - 4i F_{\mn}^{adj.} D^2 - 4 F_{\mu \rho}^{adj.} F_{\rho \nu}^{adj.} \\ \label{eq: Delta}  \hspace{2cm} + 4i (D_{\rho} F_{\rho\nu})^{adj.} D_{\mu} - 2i \delta_{\mn} (D_{\sigma} F_{\sigma\rho})^{adj.} D_{\rho} \big)\ .
\end{gather}
Evaluating (\ref{eq: 1PI 2}) requires us to split $\Delta$ into
\begin{equation}
    \Delta = {\cal P} + \tilde\Delta \ ,
\end{equation}
where $\cal{P}$ contains all the free, pure derivative terms in $\Delta$, and is therefore independent of $A_M$. This means that
\begin{equation}
    \ln\Delta = \ln {\cal P} + \ln\brac{1 + {\cal P}^{-1} \tilde{\Delta}} \ .
\end{equation}
The first term gives a field-independent constant that we can ignore, and we can evaluate the second using a power-series expansion in ${\cal P}^{-1} \tilde{\Delta}$. Explicitly, we have
\begin{equation}
    \cal{P} = \begin{pmatrix}
    - \partial^2 & \partial_t \partial_{\nu} \\
    \partial_t \partial_{\mu} & \delta_{\mn}\brac{-\partial_t^2 + \lambda^2 \partial^4}
    \end{pmatrix}\ ,
\end{equation}
with inverse (expressed in its Fourier transformed form)
\begin{equation}
    \Tilde{\cal{P}}^{-1} = \begin{pmatrix}
    \frac{\omega^2 + \lambda^2 k^4}{\lambda^2 k^6} & \frac{\omega k_{\nu}}{\lambda^2 k^6} \\
    \frac{\omega k_{\mu}}{\lambda^2 k^6} & \inv{\omega^2 + \lambda^2 k^4} \brac{\delta_{\mn} + \frac{\omega^2 k_{\mu} k_{\nu}}{\lambda^2 k^6}}
    \end{pmatrix} \ .
\end{equation}

\subsection{Renormalization of \texorpdfstring{$g$}{g} in the Pure Gauge Theory}

The complexity of (\ref{eq: Delta}) makes a full calculation of (\ref{eq: 1PI 2}) challenging. We will be slightly less ambitious and limit ourselves to finding the 1-loop quantum correction to the gauge coupling $g$. This is made much easier if we exploit the background field gauge invariance of $\Gamma[A]$. Let us restrict attention to the terms in (\ref{eq: 1PI 2}) that are quadratic in $A_t$. As $A_t$ can only enter $\Gamma[A]$ as part of either $F_{t\mu}$ or $D_t$, a contributing term of the form
\begin{equation}
    \frac{\cal{A}}{2} \fint \Tr\brac{ \partial_{\mu} A_t \partial_{\mu} A_t}\ ,
\end{equation}
must enter $\Gamma[A]$ as part of the term
\begin{equation}
    \Gamma[A] - S[A] = \frac{\cal{A}}{2} \fint \Tr\brac{F_{t\mu} F_{t\mu} } + ... \ .
\end{equation}
From this, we can read off the relation
\begin{equation}
    \inv{g_{\eff}^2} = \inv{g^2} + {\cal A}\ ,
\end{equation}
that defines the effective coupling. This is a considerable simplification: the full 1-loop correction to $g$ can be calculated by restricting our attention to terms quadratic in $A_t$. We can formally set $A_{\mu} = 0$ for the purposes of this calculation, for which $\Tilde{\Delta}$ becomes
\begin{gather} \nonumber
    \Tilde{\Delta}_{tt} = 0 \\ \nonumber
    \Tilde{\Delta}_{t\mu} = -i \brac{ A_t^{adj.} \partial_{\mu} + 2 \brac{\partial_{\mu} A_t^{adj.}}} \\ \nonumber
    \Tilde{\Delta}_{\mu t} = -i \brac{ A_t^{adj.} \partial_{\mu} - \brac{\partial_{\mu} A_t^{adj.}}} \\
    \Tilde{\Delta}_{\mn} = i \delta_{\mn} \brac{ 2 A_t^{adj.} \partial_t + \brac{\partial_t A_t^{adj.}} - i A_t^{adj.} A_t^{adj.}} \ .
\end{gather}
Here we have introduced a basis $T^a$ for our Lie-algebra with
\begin{align}
	{\rm Tr}(T^aT^b) = \frac12 \delta^{ab} \ ,
\end{align}
in terms of which the adjoint action is
\begin{align}
    \tensor{\big(A_M^{adj.}\big)}{^a_b} X^b = i \tensor{f}{^a_{bc}} A_M^b X^c \ .
\end{align}

Our final result is
\begin{equation}
    \Gamma[A] - S[A] = - \frac{3 C_2(G)}{4 (4\pi)^2 \lambda} \ln\brac{\frac{\mu^2}{\Lambda^2}} \fint \Tr ( \partial_{\mu} A_t \partial_{\mu} A_t) + ...
\end{equation}
where $\Lambda$ is an arbitrary scale introduced to isolate the local part of the expression. The effective coupling is then
\begin{equation}
    \inv{g_{\eff}^2} = \inv{g(\mu)^2} -\frac{3}{2} \frac{C_2(G)}{(4\pi)^2\lambda} \ln \brac{\frac{ \mu^2}{ \Lambda^2}} \ ,
\end{equation}
so we have the 1-loop beta function
\begin{equation}
    \beta(g) = \mu \frac{d g(\mu)}{d\mu} = -\frac{3}{2} \frac{ C_2(G) g^3}{(4\pi)^2 \lambda} < 0 \ .
\end{equation}
This in agreement with the result reported in \cite{Horava:2008jf} (which itself is based on calculations in \cite{Namiki:1992wf,Bern:1986jj,Okano:1986vr}).

It is beyond the scope of this paper to compute the $\beta$-function for $\lambda$. However, we quote the results obtained in \cite{Horava:2008jf}:\footnote{Note that \cite{Horava:2008jf} uses $e_{H} = g$ and $g_{H} = g/\lambda $ and obtains $\beta(e_{H}) = -\tfrac32 C_2(G) e_{H}^2g_{H}$ and $\beta(g_{H})= -\tfrac{35}{6}C_2(G)e_{H}g^2_{H}$ which we have converted to the above expressions for $\beta(g)$ and $\beta(\lambda)$.}
\begin{align}
	 \beta(\lambda) = \mu \frac{d \lambda(\mu)}{d\mu} =  \frac{13}{3}\frac{ C_2(G) g^2}{(4\pi)^2}\ .
\end{align}

\subsection{Bosonic Matter Contributions}

It will be interesting to see how the beta function of our coupling varies as we add matter to the theory. The simplest field we can consider is a complex scalar, which we'll denote by $X$, in a representation $R$ of $\frak{g}$. The most general minimally-coupled Lifshitz-invariant action without scalar self-interactions is
\begin{equation}
    S_{X}[\varphi, A] = \fint \brac{ \abs{D_t X}^2  + \lambda_{\varphi}^2 \abs{D^2X}^2 } \ ,
\end{equation}
where
\begin{equation}
    D_M X = \partial_M X - i A_M^{(R)} X\ ,
\end{equation}
and $\abs{X}^2 \equiv X^{\dag} X$ is a Hermitian inner product on the representation space of $R$. Note that we have differentiated $\lambda_{X}$ from the analogous quantity defined for the gauge field; though we can choose them to coincide at a specific point along the RG flow, the beta functions of the couplings will generically be different \cite{Iengo:2009ix} so such an identification is not possible in general. 
 Of course, there may still be other interaction-like terms but if these preserve the Lifstitz scaling then they won't involve the gauge field and hence won't contribute a lowest order to $\beta_g$ and $\beta_\lambda$.

To make contact with the previous section, we should expand this action about a background field $A_M$. However, as there is no overall factor of $g$ in $S_{X}$ we find
\begin{equation}
    S_{X}[X, A + g a] = S_{X}[X, A] + O(g) \ ,
\end{equation}
so to 1-loop we are able to ignore all effects of the fluctuating gauge field. In this approximation, the partition function of $X$ is
\begin{align} \nonumber
    \cal{Z}_{X}[A] &= \int DX\, e^{-S_{X}[X, A]} \\
    &= \inv{\det \brac{ - D_t^2 + \lambda_{X}^2 D^4}}\ ,
\end{align}
which changes the 1PI action to
\begin{align}
    \Gamma_X[A] &= \Gamma[A] - \ln\cal{Z}_X[A] \\ \nonumber
    &= \Gamma[A] + \Tr\ln\brac{-D_t^2  + \lambda_{X}^2 D^4 } \ .
\end{align}
As above, we want to isolate the terms quadratic in $A_t$ appearing in this expression. Setting $A_{\mu} = 0$, we find
\begin{gather}  \nonumber
    \Tr\ln\brac{- D_t^2+ \lambda_{X}^2 D^4 } = \Tr \ln {\cal P} + \Tr \ln \brac{1 + {\cal P}^{-1} \Delta} \ , \\ \nonumber
    {\cal P} = - \partial_t^2 + \lambda_{X}^2 \partial^4  \\ \label{eq: scalar contribution}
    \Delta = i \brac{2 A_t^{(R)} \partial_t + \brac{\partial_t A_t^{(R)}}} + A_t^{(R)} A_{t}^{(R)} \ .
\end{gather}
We can then expand the second term in powers of $A_t$; this gives
\begin{gather} \nonumber
    \Tr \ln \brac{1 + {\cal P}^{-1} \Delta} = T(R) \mu^{4-d}  \int \frac{d^d p dE}{(2\pi)^{d+1}} \, \Tr\brac{\Tilde{A}_t(-p,-E) \Tilde{A}_t(p,E)} \,\cal{I} + O(A_t^3) \ , \\ 
    {\cal I} = \frac{\lambda_X}{2}\int \frac{d^d k}{(2\pi)^d} \, \frac{1}{ k^2}\frac{1}{(k+p)^2}\frac{(k^2+(k+p)^2) (k^2 - (k+p)^2)^2}{E^2 + \lambda_X^2(k^2 + (p+k)^2)^2} \ . \end{gather}
The contribution to the running of $g$ is, by gauge invariance, independent of $E$, so we can set $E=0$ in the integrand. Computing the resulting integral gives
\begin{equation}
    \Tr \ln \brac{1 + {\cal P}^{-1} \Delta} = \frac{T(R) \Gamma\brac{\frac{\epsilon}{2}}}{4 (4\pi)^2 \lambda_{X}} \int \frac{d^d p dE}{(2\pi)^{d+1}}\, \Tr\brac{ \Tilde{A}_t(-p,-E) \Tilde{A}(p,E)} \brac{\frac{4\pi \mu^2}{p^2}}^{\frac{\epsilon}{2}} + O(A_t^3) \ .
\end{equation}
We will use modified minimal subtraction to remove the divergence and isolate the local part of the resulting expression, as above. We find
\begin{equation}
    \Gamma_X[A] - \Gamma[A] = \frac{T(R)}{4 (4\pi)^2 \lambda_{X}} \ln \brac{\frac{\mu^2}{\Lambda^2}}  \fint \Tr\brac{\partial_{\mu} A_t \partial_{\mu} A_t} + ...\ ,
\end{equation}
so a scalar field contributes to $\beta_g$ with a positive sign. Naturally, if we consider a real scalar field (in a real representation) then we obtain half the result given above.

We can also consider the 1-loop contribution of such a scalar field to $\beta_\lambda$. We can read this off by considering the terms quadratic in $A_{\mu}$, so we are free to set $A_t = 0$. The field-dependent terms in \eqref{eq: scalar contribution} take the form
\begin{align} \nonumber
    \Delta = - i \lambda_X^2 \bigg[ 2 \Big(2 A^{R}\cdot \partial + \brac{\partial \cdot A^{R}}\Big) \partial^2 &
    + 2 \Big(2 \brac{\partial_{\mu} A_{\nu}^{R}}\partial_{\nu} + \brac{\partial_{\mu} \partial\cdot A^{R}}\Big) \partial_{\mu} \\
    & +  2 \brac{\partial^2 A_{\mu}^{R}} \partial_{\mu} + \brac{\partial^2 \partial\cdot A^{R}} \bigg] + O(A_{\mu}^3) \ ,
\end{align}
and expanding in powers of $A_{\mu}$ gives
\begin{gather} \nonumber
    \Tr \ln \brac{1 + {\cal P}^{-1} \Delta} = T(R) \mu^{4-d} \int \frac{d^d p dE}{(2\pi)^{d+1}} \, \Tr\brac{\Tilde{A}_{\mu}(-p,-E) \Tilde{A}_{\nu}(p,E)} \,\cal{I}_{\mn} + O(A_{\mu}^3) \ , \\
    \cal{I}_{\mn} = - \frac{\lambda_X^4}{2} \int \frac{d^d k d\omega}{(2\pi)^{d+1}} \, \frac{(2k+p)_{\mu} (2k + p)_{\nu} (k^2 + (k+p)^2)^2}{(\omega^2 + \lambda_X^2 k^4)( (\omega + E)^2 + \lambda_X^2 (k+p)^4)} \ .
\end{gather}
As we are only interested in the spatial derivatives of the gauge field, we can again set $E=0$ in the integrand. The integral over $\omega$ leaves us with the momentum integral
\begin{equation}
    \cal{I}_{\mn} = - \frac{\lambda_X}{4}\int \frac{d^d k}{(2\pi)^d}\, (2k+p)_{\mu} (2k+p)_{\nu} \brac{\inv{(k+p)^2} + \inv{k^2}} \ ,
\end{equation}
which vanishes in dimensional regularisation. The scalar field then gives no contribution to the running of $\lambda$ at 1-loop.

\subsection{Fermionic Matter Contributions}

Next, let us consider adding an anticommuting complex scalar field $\psi$ in the representation $R$ to the theory. We will take $\psi$ to have the action
\begin{equation}
    S_{\psi} = \fint \brac{\psi^{\dag} D_t \psi + \lambda_{\psi} D_{\mu}\psi^{\dag} D_{\mu} \psi } \ .
\end{equation}
Following the same steps as for the commuting scalar, the contribution to the 1PI effective action is
\begin{equation}
    \Gamma_{\psi}[A] - \Gamma[A] = - \Tr \ln \brac{D_t - \lambda_{\psi} D^2 } \ .
\end{equation}
Let us first consider the contribution to the running of $g$. Setting $A_{\mu} = 0$ and expanding this gives, ignoring a field-independent constant,
\begin{gather} \nonumber
    \Tr\ln(D_t - \lambda_{\psi} D^2 ) = - T(R) \mu^{4-d} \int \frac{d^d p dE}{(2\pi)^{d+1}} \, \Tr\brac{\Tilde{A}_t(-p,-E) \Tilde{A}_t(p,E)} \, {\cal I} + O(A_t^3) \ , \\
    {\cal I} = \int \frac{d^d k d\omega}{(2\pi)^{4+1}} \, \frac{(i\omega + \lambda_{\psi} k^2) (i (\omega + E) + \lambda_{\psi} (k+p)^2)}{(\omega^2 + \lambda_{\psi}^2 k^4)((\omega + E)^2 + \lambda^2_{\psi} (k+p)^4}\ .
\end{gather}
Evaluating the integral over $\omega$ shows that this term vanishes, and so anticommuting scalars will not contribute to the running of $g$. The same arguments apply to terms involving the spatial gauge fields and two propagators; noting that in dimensional regularisation any integral over a single scaleless propagator is automatically zero, we find that the terms quadratic in $A_{\mu}$ also vanish and anticommuting scalars do not contribute to the running of $\lambda$. This is in line with the non-renormalization theorem observed in \cite{Arav:2019tqm}. As the components of the spinor field in a Lifshitz-invariant minimally-coupled spinor action decouple from each other, these results extend to spinorial Fermions. However, if we were to consider a spinor field with further interactions or a relevant deformation this conclusion would not hold.

In summary, we find that the 1-loop $\beta$-functions, including both Bosonic and Fermionic scalar contributions, are
\begin{align}
	\beta(g) &=  -\frac{3}{2} \frac{ C_2(G) g^3}{(4\pi)^2 \lambda} + \frac{T_{scalar}(R)}{2(4\pi)^2 \lambda_{X}}\nonumber\\
\beta(\lambda) &=	\frac{13}{3}\frac{ C_2(G) g^2}{(4\pi)^2}\ ,\end{align}
where $T_{scalar}(R)$ comes from the complex Bosonic scalars.

\section{No-go Theorems and Resolutions}

\subsection{Spinorial Supersymmetry} \label{sect: no-go susy}

We would like to augment our Bosonic symmetry algebra with spinorial supercharges. The minimal consistent addition is a single $SO(4)$ Weyl spinor characterised by the brackets
\begin{align} \nonumber
    [M_{\mn}, Q_{\alpha}] &= \frac{i}{2} \tensor{\brac{\gamma_{\mn}}}{_\alpha^\beta} Q_{\beta} \ , \\ \nonumber
    [T, Q_{\alpha}] &= i Q_{\alpha} \ , \\
    \{Q_{\alpha}, Q^{\dag \beta}\} &= 2 \delta_{\alpha}^{\beta} H \ .
\end{align}
As a larger superalgebra with the same Bosonic symmetries will necessarily contain this as a subalgebra, any obstructions to realising this algebra in a field theory will extend to the more general case. Defining the action of $Q_{\alpha}$ on a general field $\Phi$ as
\begin{equation}
    \delta_{\xi} \Phi = (\xi^{\dag \alpha} Q_{\alpha} + Q^{\dag \alpha} \xi_{\alpha}) \Phi \ ,
\end{equation}
we see that the requirement that the supercharges close onto $H$ becomes the commutator
\begin{equation} \label{eq: susy commutator}
    [\delta_{\xi}, \delta_{\eta}] \Phi = 2i (\xi^{\dag} \eta - \eta^{\dag} \xi) \partial_t \Phi\ ,
\end{equation}
on the fields.

Let us attempt to formulate a supersymmetric extension of a single complex Lifshitz scalar field $X$ with action
\begin{equation}
    S_{X} = \fint \brac{\partial_t \bar{X} \partial_t X - \lambda^2 \partial^2 \bar{X} \partial^2 X} \ .
\end{equation}
Both $X$ and $Q_{\alpha}$ have scaling dimension 1, so to get consistent transformations of our fields under $Q_{\alpha}$ we require that any Fermions we add to the action must have integer dimension. We will denote our Fermion species by $\{\psi_i\}$; since we want a free field theory with no dimensionful parameters, the most general Fermion action we can write down is
\begin{equation} \label{eq: Fermion action}
    S_{F} = \sum_{i,j} \brac{ \fint \psi^{\dag}_i \cal{L}_{ij} \psi_j + h.c. }\ ,
\end{equation}
in terms of the differential operators $\{\cal{L}_{ij}\}$ of dimension $[\cal{L}_{ij}] = 6 - [\psi_i] - [\psi_j] \geq 0$. 

Suppose that we have a Fermion $\psi_i$ for which $[\psi_i] \geq 3$. The lack of dimensionful parameters means that $\psi_i$ cannot appear in the transformation of $X$. However, $\delta S_F$ will necessarily include terms proportional to $\psi_i$ of the form
\begin{equation}
    \delta S_F \supset \sum_j \brac{\fint \psi_i^{\dag} \cal{L}_{ij} \delta\psi_j + h.c. } \ .
\end{equation}
This means there are terms in $\delta S_F$ that we cannot cancel with terms in $\delta S_X$ unless $\delta \psi_j$ = 0 for all fields that couple to $\psi_i$. This is only consistent with the closure of the algebra on the field if the equation of motion for $\psi_j$ is
\begin{equation}
    \partial_t \psi_j = 0 \ .
\end{equation}
As the scalar field $X$ is dynamical, it would be inconsistent for $\psi_j$ to appear in its transformation. We see that Fermions of dimension greater than 2 and the lower-dimension Fermions they couple to will not affect the supersymmetry properties of the theory, so we are free to restrict ourselves to fields of dimension $[\psi_i] \leq 2$.

Unitarity forces us to consider at most one time-derivative in each term in the Fermion action. We can then parameterise \eqref{eq: Fermion action} as
\begin{equation}
    S_F = \sum_{i,j} \bigg[ \fint \brac{ \alpha_{ij} \psi^{\dag}_i \partial_t \slashed{\partial}^{4 - [\psi_i] - [\psi_j]} \psi_j + \beta_{ij} \psi^{\dag}_i \slashed{\partial}^{6 - [\psi_i] - [\psi_j]} \psi_j} + h.c. \bigg] \ ,
\end{equation}
which can be conveniently rewritten in terms of the redefined fields
\begin{equation}
    \psi'_i = \slashed{\partial}^{2 - [\psi_i]} \psi_i\ ,
\end{equation}
as
\begin{equation}
    S_F = \sum_{i,j} \bigg[ \fint \brac{ \alpha_{ij} \psi^{\prime\dag}_i \partial_t \psi^{\prime}_j + \beta_{ij} \psi^{\prime\dag}_i \partial^2 \psi^{\prime}_j} \bigg] \ .
\end{equation}

The lack of dimensionful parameters constrains the allowed supersymmetry transformations to be of the form
\begin{align} \nonumber
    \delta X &= \xi^{\dag}\sum_i \sigma_i \slashed{\partial}^{2 - [\psi_i]} \psi_i \\
    \delta \psi_i &= (L_iX) \xi\ ,
\end{align}
for some set of differential operators $\{L_i\}$. At first it appears that we could be more general and include time-derivatives in $\delta X$. However, to recover \eqref{eq: susy commutator} we would be forced to exclude time-derivatives from the operators $\{L_i\}$, which would imply the equation of motion for $X$ is first-order in time. As the commutator of the supercharge with the momentum or Hamiltonian vanishes, these can be written in terms of the redefined fields as
\begin{align} \nonumber
    \delta \varphi &= \xi^{\dag}\sum_i \sigma_i \psi'_i \\
    \delta \psi'_i &= \slashed{\partial}^{2 - [\psi_i]}(L_i X) \xi \equiv (L'_i X) \xi \ .
\end{align}
We see that we lose nothing by only considering dimension 2 Fermions, which we shall do from this point onwards.

Matching the number of Bosonic and Fermionic degrees of freedom suggests we take a single $SO(4)$ Weyl spinor $\psi$ as the superpartner of $X$; the only possible free Lifshitz-invariant action, scaling $\psi$ to have a canonical kinetic term and fixing a constant, is
\begin{equation}
    S = \fint \brac{ \partial_t \Bar{} \partial_t X- \lambda^2 \partial^2 \Bar{X} \partial^2 X + i \psi^{\dag} \partial_t \psi - \lambda \partial_{\mu} \psi^{\dag} \partial_{\mu} \psi} \ .
\end{equation}
The supersymmetry transformations consistent with Lifshitz scaling symmetry are
\begin{align} \nonumber
    \delta X &= \xi^{\dag} \psi \\ 
    \delta \psi &= \alpha \partial_t X \, \xi + \beta \partial^2 X\, \xi\ ,
\end{align}
which means \eqref{eq: susy commutator} is only satisfied if $\beta = 0$. However, the transformation of the action is
\begin{equation}
    \delta S = \fint \Big( \partial_t \psi^{\dag} \partial_t X (1 - i \alpha) - \partial^2 \psi^{\dag} \partial^2 X \brac{\lambda^2 - \lambda \beta} + \psi^{\dag} \partial_t \partial^2 \brac{i \beta + \lambda \alpha}\Big) \xi + h.c.\ ,
\end{equation}
so we only have $\delta S=0$ if $\beta = i \lambda \alpha$ and $\alpha = - i$. The procedure is then inconsistent and we conclude that the free complex Lifshitz scalar does not admit a linearly realised spinorial symmetry if Lifshitz scaling symmetry is preserved.

It is interesting to consider how this argument changes if we instead consider a real scalar field $\varphi$. A brief calculation shows that the action
\begin{equation}
    S = \fint \brac{\inv{2} \partial_t \varphi \partial_t \varphi - \frac{\lambda^2}{2} \partial^2 \varphi \partial^2 \varphi + i \psi^{\dag} \partial_t \psi - \lambda \partial_{\mu} \psi^{\dag} \partial_{\mu} \psi}\ ,
\end{equation}
is invariant under the transformations
\begin{align} \nonumber
    \delta \varphi &= \xi^{\dag} \psi + \psi^{\dag} \xi \\
    \delta \psi &= -i \partial_t \varphi \xi + \lambda \partial^2 \varphi \xi \ .
\end{align}
We find that the commutator of the transformations on $\varphi$ is
\begin{equation}
    [\delta_{\xi}, \delta_{\eta}] \varphi = 2i (\xi^{\dag} \eta - \eta^{\dag} \xi) \partial_t \varphi,
\end{equation}
as required by \eqref{eq: susy commutator}, so this appears to be a valid representation of the superalgebra on the fields. However, the number of Bosonic and Fermionic degrees of freedom no longer match; we would therefore not expect this theory to admit an interacting extension preserving the supersymmetry. We may hope that we could add more Bosonic degrees of freedom to rectify this, but the addition of a second scalar field just takes us back to the complex scalar that we have already considered and ruled out. If we only wish to consider theories with four spatial dimensions, we can then also rule out interacting supersymmetric extensions of the real Lifshitz scalar field. However, suppose we allow ourselves to instead work in two spatial dimensions. Our Fermions will then lie in spinorial representations of $SO(2)$; the Weyl representations of this group are one-dimensional, so our field $\psi$ has a single complex Grassmann-valued component. The number of on-shell Bosonic and Fermionic degrees of freedom now match, and there is no barrier to the existence of interacting extensions. Theories of this sort were constructed in \cite{Chapman:2015wha}.

Let us now analyse the same problem for an Abelian Lifshitz gauge theory. We choose to work in Weyl gauge, $A_t = 0$, and from the discussion of section \ref{sect: field equations} we can also set $\partial_{\mu} A_{\mu}$ to zero. With this choice, the gauge field's action becomes
\begin{equation}
    S_G = \inv{2} \fint \brac{\partial_t A_{\mu} \partial_t A_{\mu} - \lambda^2 \partial^2 A_{\mu} \partial^2 A_{\mu} } \ .
\end{equation}
This contains three independent propagating degrees of freedom, so to match Bosonic and Fermionic excitations we require two Weyl spinors and an additional real scalar. The physical Bosonic degrees of freedom associated with the gauge field have dimension 1, so as above we only need to consider Fermions of dimension 2. Since we are working with a free theory, the scalars will decouple from the gauge field and can be ignored. The existence of two independent Weyl spinors that are related by supersymmetry to the same gauge field requires us to work with a superalgebra containing two Weyl supercharges. However, in the free theory we can choose to focus on the subalgebra containing a single Weyl supercharge and hence with a single Fermion. Any barriers we find to a consistent realisation of this subalgebra on the fields will then extend to the full theory.

The consistent supersymmetry variations of the gauge and Weyl spinor fields are
\begin{align} \nonumber
    \delta A_{\mu} &= i \xi^{\dag} \gamma_{\mu} \psi - i \psi^{\dag} \gamma_{\mu} \xi \\
    \delta \psi &= \alpha \gamma_{\mu} \partial_t A_{\mu} \xi + \beta \gamma_{\mu} \partial^2 A_{\mu} \xi\ ,
\end{align}
where we have taken the Fermion and supercharge to have opposite chiralities. The variation of the action is
\begin{align}
    \delta S = \fint \xi^{\dag} \gamma_{\mu} \Big[ i (1 + \bar{\alpha}) \partial_t \psi \partial_t A_{\mu} - i \lambda (\lambda + i \bar{\beta}) \partial^2 \psi \partial^2 A_{\mu} + (i \bar{\beta} - \lambda \alpha) \partial^2 A_{\mu} \partial_t \psi \Big] + h.c.\ ,
\end{align}
which only vanishes if we take $\alpha = -1$ and $\beta = -i \lambda$. However, with these parameters we have
\begin{equation}
    [\delta_{\xi}, \delta_{\eta}] A_{\mu} = 2i\brac{\xi^{\dag} \eta - \eta^{\dag} \xi} \partial_t A_{\mu} + 2\lambda \brac{ \xi^{\dag} \gamma_{\mn} \eta - \eta^{\dag} \gamma_{\mn} \xi} \partial^2 A_{\nu}\ ,
\end{equation}
which does not close onto time translations. Like the scalar case, an Abelian Lifshitz gauge theory does not admit a spinorial supersymmetric extension. As the Abelian theory can be seen as the $g\to0$ limit of the non-Abelian theory, we can extend this conclusion to the full interacting theory.

\subsection{Boost Symmetry}

We will consider the behaviour of the pure gauge theory under the Galilean boost
\begin{align} \nonumber
    t &\to t' = t \\
    x^{\mu} &\to x^{\prime\mu} = x^{\mu} + v^{\mu} t \ .
\end{align}
Requiring that our field transforms in a way consistent with gauge transformations forces us to take
\begin{align} \nonumber
    A'_{\mu}(t',x') &= A_{\mu}(t,x) \\
    A'_t(t',x') &= A_t(t,x) - v^{\mu} A_{\mu}(t, x) \ ,
\end{align}
so the new field strength components are
\begin{align} \nonumber
    F'_{\mn}(t',x') &= F_{\mn}(t,x) \\
    F'_{t\mu}(t',x') &= F_{t\mu}(t,x) - v^{\nu} F_{\nu\mu}(t,x) \ .
\end{align}
Using these, the transformation of the gauge action is
\begin{equation}
    S_g' = S_g + \inv{2g^2} \fint \brac{2 v^{\nu} \Tr\brac{F_{t\mu} F_{\mn}} + { O}(v^2)}\ ,
\end{equation}
so this is not a symmetry of the theory. 

We can do better by introducing new fields into the theory. This mirrors the way boost symmetry is introduced in Galilean electrodynamics \cite{LeBellac:1972, Santos_2004, Festuccia:2016caf}, though our intention will be to find a different  boost-invariant action that preserves  the kinetic term for the gauge field. For convenience, we will work with infinitesimal $v^\mu$. We then have two ways to render the theory boost-invariant. We can either introduce a vector field $B_{\mu}$ into the action with the coupling and boost transformation
\begin{align} \nonumber
    \Delta S_X &= \inv{2g^2}  \fint \Tr\brac{ B_{\mu} F_{t\mu} } \ ,\\
    B'_{\mu}(t',x') &= B_{\mu}(t,x) - 2 v^{\nu} F_{\mn}(t,x) + { O}(v^2) \ ,
\end{align}
or a 2-form field $G_{\mn}$ with
\begin{align} \nonumber 
    \Delta S_G &= \inv{2g^2} \fint \Tr \brac{G_{\mn} F_{\mn}} \ , \\ \label{eq: G boost transformation}
    G'_{\mn}(t',x') &= G_{\mn}(t,x) + 2 v_{[\mu} F_{|t|\nu]}(t,x) + {O}(v^2) \ .
\end{align}
Since we can't add further terms involving $B_{\mu}$ or $G_{\mn}$ without spoiling the symmetry, the fields act as Lagrange multipliers. As we want the gauge field's kinetic term to be retained, we discount the first case. This leaves us with the boost-invariant action
\begin{equation}
    \tilde{S} = \inv{2g^2}  \fint \Tr \brac{ F_{t\mu} F_{t\mu} + G_{\mn} F_{\mn} - \lambda^2 D_{\nu} F_{\nu\mu} D_{\rho} F_{\rho\mu}} \ .
\end{equation}
However, the constraint imposed by $G_{\mn}$ sets $F_{\mn} = 0$; this kills off the final term, so the above action $\tilde{S}$ is equivalent to
\begin{equation}
    S = \inv{2g^2}  \fint \Tr \brac{F_{t\mu} F_{t\mu} + G_{\mn} F_{\mn}} \ .
\end{equation}

As it stands, the theory above is trivial: the Lagrange multiplier forces the spatial components of the gauge field to satisfy $F_{\mn} = 0$, giving us field configurations that gauge-equivalent to $A_\mu  = 0$. However, this can be rectified by deforming the theory in a way that preserves the boost symmetry and allows for non-trivial field configurations, as we shall see momentarily.

\subsection{Supersymmetry Revisited}

The main barrier to the existence of supersymmetric Lifshitz actions was the higher spatial derivative term. Since the Lagrange multiplier kills off terms of this form, we may wonder if the constrained actions can be made supersymmetric. A short calculation shows that the free theory
\begin{equation}
    S = \fint \brac{ \inv{2} F_{t\mu} F_{t\mu} + \inv{2} G_{\mn} F_{\mn} + i \psi^{\dag} \partial_t \psi + i \chi^{\dag} \gamma_{\mu} \partial_{\mu} \psi + i \psi^{\dag} \gamma_{\mu} \partial_{\mu} \chi}\ ,
\end{equation}
has vanishing variation under the transformation
\begin{align} \nonumber
    \delta A_{\mu} &= i \xi^{\dag} \gamma_{\mu} \psi - i \psi^{\dag} \gamma_{\mu} \xi \\ \nonumber
    \delta A_t &= -i \xi^{\dag} \chi + i \chi^{\dag} \xi \\ \nonumber
    \delta G_{\mn} &= - i \xi^{\dag} \gamma_{\mn} \partial_t \chi - i \partial_t \chi^{\dag} \gamma_{\mn} \xi \\ \nonumber
    \delta \psi &= - \gamma_{\mu} F_{t\mu} \xi \\
    \delta \chi &= - \inv{4} \gamma_{\mn} G_{\mn} \xi\ ,
\end{align}
if we take $G_{\mn}$ to be either self-dual or anti-self-dual, subject to the chirality condition
\begin{equation} \label{eq: chirality condition}
    \inv{2}\epsilon_{\mn\rho\sigma} G_{\rho\sigma} \gamma_* \psi = G_{\mn} \psi \ .
\end{equation}

For definiteness we shall take $\gamma_* \psi = -\psi$ and $G_{\mn}$ anti-self-dual, which forces us to take both $\gamma_* Q = Q$ and $\gamma_* \chi = \chi$. Recalling that (working in the chiral basis for convenience) the $SO(4)$ Clifford algebra matrices $\gamma_{\mn}$ have the chiral decomposition
\begin{equation}
    \gamma_{\mn} = \begin{pmatrix} \sigma_{\mn} & 0 \\
    0 & \bar{\sigma}_{\mn}
    \end{pmatrix}\ ,
\end{equation}
with $\sigma_{\mn}$ anti-self-dual and $\bar{\sigma}_{\mn}$ self-dual, we see that the transformation of $G_{\mn}$ takes the form
\begin{equation}
    \delta G_{\mn} = i \xi^{\dag} \sigma_{\mn} \partial_t \chi + i \partial_t \chi^{\dag} \sigma_{\mn} \xi\ .
\end{equation}
Thankfully, this is consistent with the requirement that $G_{\mn} + \delta G_{\mn}$ remains anti-self-dual. A brief calculation shows that the gauge-covariant generalisation of \eqref{eq: susy commutator} is satisfied for the Bosonic fields, with
\begin{align} \nonumber
    [\delta_{\xi}, \delta_{\eta}] A_{\mu} &= 2i(\xi^{\dag} \eta - \eta^{\dag} \xi ) F_{t\mu} \\ \nonumber
    [\delta_{\xi}, \delta_{\eta}] A_{t} &= 0 \\
    [\delta_{\xi}, \delta_{\eta}] F_{\mn} &= 2i(\xi^{\dag} \eta - \eta^{\dag} \xi ) \partial_t G_{\mn} \ ,
\end{align}
so we may hope that interacting extensions can be formed. Theories of this sort are known and have the gauge-sector action
\begin{align} \nonumber
    S = \inv{2g^2} \fint \Tr\Big( F_{t\mu} F_{t\mu} & + G_{\mn} F_{\mn} -  D_{\mu} \phi^{\dag} D_{\mu} \phi + i \psi^{\dag} D_t \psi \\ \label{eq: full action}
    & + i \chi^{\dag} \gamma_{\mu} D_{\mu} \psi + i \psi^{\dag} \gamma_{\mu} D_{\mu} \chi + i \psi^{\dag} [\phi, \psi] \Big) \ ,
\end{align}
where we've included an additional real scalar field $\phi$. This action, possibly dressed-up to include additional scalars and Fermions in other representations of the gauge group, is in the class of actions that were previously constructed by considering null reductions of six-dimensional superconformal field theories and as such admit supersymmetry. We will not give the details here but refer the interested reader to \cite{Lambert:2018lgt,Lambert:2020jjm}. 
%On the other hand the pure gauge sector of these theories is similar to the  action of \cite{Chalmers:} for self-dual gauge fields.

Though giving $G_{\mn}$ a self-duality condition allows us to write down a supersymmetric extension of the theory, it naively appears to break invariance under boosts as the transformation \eqref{eq: G boost transformation} does not leave $G_{\mn}$ anti-self-dual. We can rectify this by proposing that the fields in \eqref{eq: full action} have the infinitesimal transformations
\begin{align} \nonumber
    A'_{\mu}(t',x') &= A_{\mu}(t,x) \\ \nonumber
    A'_t(t',x') &= A_t(t,x) - v^{\mu} A_{\mu}(t,x) \\ \nonumber
    G'_{\mn}(t',x') &= G_{\mn}(t,x) + 2 v_{[\mu} F_{|t|\nu]}(t,x) \pm \epsilon_{\mn\rho\sigma} v_{\rho} F_{t\sigma} (t,x) \\ \nonumber
    \phi'(t',x') &= \phi(t,x) \\ \nonumber
    \psi'(t',x') &= \psi(t,x) \\
    \chi'(t',x') &= \chi(t,x) + \inv{2} v^{\mu} \gamma_{\mu} \psi(t,x)\ ,
\end{align}
under boosts. The variation of the action is then
\begin{equation}
    S' = S \pm \frac{v^{\mu}}{2g^2} \fint \epsilon_{\mn\rho\sigma} \Tr\brac{ F_{t\nu} F_{\rho\sigma}} + O(v^2) \ ,
\end{equation}
which still appears to not vanish. However, it can be shown that this term is a total derivative: we therefore find that the action is both boost-invariant and supersymmetric, as hoped.

If we also consider the infinitesimal coordinate transformation
\begin{equation}
    t\to t' = t + \omega t^2 \qquad x^{\mu} \to x^{\prime \mu} = x^{\mu} + \omega t x^{\mu}\ ,
\end{equation}
for which we postulate the field transformations
\begin{align} \nonumber
    A'_{\mu}(t',x') &= \brac{1 - \omega t} A_{\mu}(t,x) \\ \nonumber
    A'_t(t',x') &= (1 - 2\omega t) A_t(t,x) - \omega x^{\mu}A_{\mu}(t,x) \\ \nonumber
    G'_{\mn}(t',x') &= (1 - 4\omega t) G_{\mn})(t,x) + 2\omega x_{[\mu} F_{|t|\nu]}(t,x) \pm \omega \epsilon_{\mn\rho\sigma} x^{\rho}  F_{t\sigma}(t,x) \\ \nonumber
    \phi'(t',x') &= (1 - 2\omega t) \phi(t,x) \\ \nonumber
    \psi'(t',x') &= (1 - 2\omega t) \psi(t,x) \\
    \chi'(t',x') &= \brac{1 - 3\omega t} \chi(t,x) + \inv{2}\omega x^{\mu} \gamma_{\mu} \psi(t,x) \ ,
\end{align}
we see the change in the action is
\begin{equation}
    S' = S \mp \frac{\omega}{2g^2} \epsilon_{\mn\rho\sigma} \fint x^{\sigma} \Tr \brac{ F_{\mn} F_{t\rho}} + O(\omega^2) \ .
\end{equation}
As above, the term linear in $\omega$ can be shown to be a total derivative so this is also a symmetry. 

Let us examine the Bosonic spacetime symmetry algebra of our theory. The symmetry transformations are generated by the Hermitian vector fields
\begin{align} \nonumber
    H &= i \partial_t \\ \nonumber
    P_{\mu} &= -i \partial_{\mu} \\ \nonumber
    M_{\mn} &= -i \brac{x_{\mu} \partial_{\nu} - x_{\nu} \partial_{\mu}} \\ \nonumber
    T &= -i (2t \partial_t + x^{\mu} \partial_{\mu}) \\ \nonumber
    B_{\mu} &= -i t \partial_{\mu} \\
    K &= -i \brac{t^2 \partial_t + t x^{\mu} \partial_{\mu} } \ ,
\end{align}
so the non-zero Lie algebra brackets are
\begin{align} \nonumber
    [M_{\mn}, M_{\rho\sigma}] &= i \brac{\delta_{\mu\rho} M_{\nu\sigma} + \delta_{\nu\sigma} M_{\mu\rho} - \delta_{\mu\sigma} M_{\nu\rho} - \delta_{\nu\rho} M_{\mu\sigma} } \\ \nonumber
    [M_{\mn} , P_{\rho}] &= i \brac{\delta_{\mu\rho} P_{\nu} - \delta_{\nu\rho} P_{\mu} } \\ \nonumber
    [M_{\mn} , B_{\rho}] &= i \brac{\delta_{\mu\rho} B_{\nu} - \delta_{\nu\rho} B_{\mu} } \\ \nonumber
    [T, H] &= 2i H \\ \nonumber
    [T, P_{\mu}] &= i P_{\mu} \\ \nonumber
    [T, B_{\mu}] &= -i B_{\mu} \\ \nonumber
    [T, K] &= -2i K \\ \nonumber
    [P_{\mu}, K] &= - i B_{\mu} \\ \nonumber
    [H, B_{\mu}] &= i P_{\mu} \\ 
    [H, K] &= i T \ .
\end{align}
We recognise this as the Schr\"odinger algebra with vanishing central extension \cite{Hagen:1972}.

\section{RG Flows}

In this section we wish to consider possible RG-flows starting from the Lifshitz scale invariant theories that we have studied above. In particular, while we have argued that such theories cannot have supersymmetry or boost symmetries, our aim is to argue that there do exist flows for which the IR fixed point theories do enjoy these symmetries. We should say from the outset that we do not claim that such flows are generic - merely that they exist. Our discussion here provides a quantum treatment parallel to the discussion in \cite{Lambert:2019nti}.

To begin with, note that we can deform $S_A$ to
\begin{align}\label{Sdeform}
 S^{(M)}_A & = \frac{1}{2g^2}  \fint \Tr\brac{ F_{t\mu} F_{t\mu} - \lambda^2 D_\nu F_{\nu\mu}D_\rho F_{\rho\mu } - M^2 F^\pm_{\mn}F^\pm_{\mn}} 	\ ,
 \end{align}
 where
 \begin{align}
 F^\pm_{\mu\nu} = \frac12 F_{\mu\nu} \pm \frac14 \varepsilon_{\mu\nu\lambda\rho}F_{\lambda\rho}	\ .
 \end{align}
 This deformation is similar in spirit to the discussion in \cite{Horava:2008jf}
except that we do not wish to flow to a Lorentzian fixed point.

We should be somewhat cautious, as
\begin{equation}
    F^{\pm}_{\mn} F^{\pm}_{\mn} = \inv{2}F_{\mn} F_{\mn} \pm \inv{4} \epsilon_{\mn\rho\sigma} F_{\mn} F_{\rho \sigma}\ ,
\end{equation}
so the term splits into a sum of dynamical and topological terms. The dynamical term will receive both perturbative and non-perturbative quantum corrections, whereas the topological term will only receive non-perturbative corrections; under an RG flow we then generically expect
\begin{equation}
    M^2\Tr \fint F^{\pm}_{\mn} F^{\pm}_{\mn} \to \fint \Tr\brac{ \frac{ \Tilde{M}^2 }{2} F_{\mn} F_{\mn}  \pm \frac{ \hat{M}_2^2 }{4} \epsilon_{\mn\rho\sigma} F_{\mn} F_{\rho\sigma}} \ ,
\end{equation}
with $\Tilde{M}^2\ne \hat{M}_2^2$.
However, this can be resolved by tuning the coefficient of our topological term in the UV action so its IR value coincides $\Tilde{M}^2$, {\it i.e.} rather than simply $M^2 F^\pm_{\mn}F^\pm_{\mn}$ in the UV we should start with
\begin{align}
 \fint \Tr\brac{ M^2 F^{\pm}_{\mn} F^{\pm}_{\mn}  \pm \frac{ \delta M^2 }{4} \epsilon_{\mn\rho\sigma} F_{\mn} F_{\rho\sigma}} \ ,
\end{align}
where $\delta M^2$ is chosen such that its IR value is $\Tilde{M}^2 - \hat{M}^2$. We will ignore these subtleties here and focus solely on the first term.

Following the discussion of \cite{Iengo:2009ix}, we expect quantum effects in our RG flow to be characterised by two scales. The first is the scale $\Lambda_s$ at which the theory becomes strongly-coupled; this is set by the beta function of $g$. The second is the scale at which quantum corrections due to the higher-derivative terms in the Lagrangian become small- this is the Lifshitz analogue of the freezing that occurs in the running of couplings in massive Lorentzian theories. By dimensional analysis this scale is given by the ratio of the couplings of the spatial gradient terms. As the relevant coupling $M$ would receive no quantum corrections if the higher-derivative term were absent, we see that if we choose $\Lambda_s$ such that
\begin{equation}
    \Lambda_s \ll \frac{M}{\lambda}\ ,
\end{equation}
we can decouple quantum corrections to the running of $M$ when the theory is still in the weakly-coupled regime and perturbation theory is still valid. The RG behaviour of the dimensionless coupling
\begin{equation}
    m = \frac{\Tilde{M}}{\mu} \ ,
\end{equation}
defined with respect to the value of $M$ below the decoupling point, will then be entirely classical in the IR. Importantly, this means that our conclusions should remain valid despite the theory becoming strongly-coupled.

  %We take the ratio of $\mu$ and $M$ to be small\footnote{We need to take $M$ to be large enough for this to kick in before our theory flows to the strongly-coupled region for this to be a sensible inference.} and the final term will dominate the expression.

Let us follow \cite{Gomis:2005pg} and introduce a Lagrange multiplier field
\begin{align}
 S^{(M)}_A & = \frac{1}{2g^2}  \fint \Tr \brac{ F_{t\mu} F_{t\mu} - \lambda^2 D_\nu F_{\nu\mu}D_\rho F_{\rho\mu } +\frac{1}{4M^2}G_{\mu\nu}G_{\mu\nu}+ G_{\mu\nu} F_{\mn}} 	\ ,
 \end{align}
	where  
	\begin{align}
 G_{\mu\nu} = \pm  	\frac12\varepsilon_{\mu\nu\lambda\rho}G_{\lambda\rho}\ .
 \end{align} 
 Integrating out $G_{\mu\nu}$ leads us back to (\ref{Sdeform}). 
 Note that we can preserve the Lifshitz  scaling symmetry by giving  
 $G_{\mu\nu}$ Lifshitz  scaling dimension 4.
 On the other hand, it is clear that in the IR limit where   where $M/\mu\to \infty$ we are simply left with 
\begin{align}
 S^{(IR)}_A & = \inv{2g^2}  \fint \Tr \brac{ F_{t\mu} F_{t\mu} - \lambda^2 D_\nu F_{\nu\mu}D_\rho F_{\rho\mu } +  G_{\mu\nu}F_{\mu\nu}}	\ .
 \end{align}
 Thus in the IR  the dynamics are reduced to solutions of
\begin{align} \label{eq: gauge constraint}
	F^\pm_{\mu\nu}=0\ ,
\end{align}
{\it i.e.} (anti-)self-dual gauge fields. 
In addition, we note that on the constraint surface the original fourth-order gradient term vanishes and can be dropped:
\begin{align}
 S^{(IR)}_A & = \frac{1}{2g^2} \fint \Tr\brac{F_{t\mu} F_{t\mu} + G_{\mu\nu}F_{\mu\nu}} 	\ .
 \end{align}

We can similarly consider deforming the action for adjoint complex scalars, obtaining the action
\begin{align}
S^{(M)}_{X} & = 	\inv{2 g^2}  \fint \Tr \brac{D_t X^\dag D_tX  - M^2D_\mu X^\dag D_\mu X  - \lambda^2_{X} D^2X^\dag D^2 X } \ .
\end{align}
Using the same arguments as above, in the infrared we are pushed onto the surface
\begin{align}
D_\mu X =0 \ ,	
\end{align}
and as a result the scalars are frozen. However, in this case we can introduce a new scalar
\begin{align}
\phi = MX	\ ,
\end{align}
with Lifshitz dimension 2. Now the action is
\begin{align}
S^{(M)}_{X} & = 	\inv{2g^2}  \fint dtd^4x \Tr \bigg( \frac1{M^2} D_t \phi^\dag D_t \phi - D_\mu \phi^\dag D_\mu \phi  - \frac{\lambda_{X}^2}{M^2} D^2 \phi^\dag D^2\phi\bigg)\ ;
\end{align}
we would then expect the IR theory to be described by
\begin{align}
S^{(IR)}_{X} & = 	- \inv{2g^2}\fint \Tr \brac{D_\mu \phi^\dag D_\mu \phi} \ ,
\end{align}
whose equation of motion does have solutions, at least around (anti-)self-dual backgrounds.

Finally, we consider the Fermion action
\begin{align}
S_{\psi}^{(M)}  = 	\inv{2g^2} \fint \Tr & \Big( i \psi^\dag  D_t \psi -  \lambda_{\psi}^2 D_\mu \psi^{\dag} D_\mu\psi - i \sigma_1 \psi^{\dag} \gamma_{\mu\nu}[F_{\mu\nu},\psi]\nonumber\\
&\qquad \qquad  + i M \psi^{\dag} \gamma^\mu D_\mu\psi +  i\sigma_2 M \psi^\dag[X,\psi ]\Big)\ ,
\end{align}
where we've introduced the dimensionless couplings $\sigma_1,\sigma_2$ to mediate interactions between our fields. In the IR we expect to find the constraint
\begin{align}
\gamma_\mu D_\mu \psi =0	\ .
\end{align}
This can be achieved by including a Fermionic Lagrange multiplier $\chi$ into the action:
\begin{align}
S^{(IR)}_\psi = \frac{1}{2g^2}  \fint \Tr & \Big(i \psi^\dag  D_t \psi  +i\chi^\dag \gamma_\mu D_\mu \psi +i\psi^\dag \gamma_\mu D_\mu \chi+ i\sigma_2 \psi^\dag[\phi,\psi ]\nonumber\\
&\qquad \qquad - \lambda_{\psi}^2 D_\mu \psi^{\dag} D_\mu\psi - i \sigma_1 \psi^{\dag} \gamma_{\mu\nu}[F_{\mu\nu},\psi] \Big) \ ,
\end{align}
where $\chi$ has scaling dimension 3. Note that on the Fermionic constraint surface we have
\begin{align}
0=\gamma_\mu\gamma_\nu D_\mu D_\nu\psi    = D^2\psi - i \gamma_{\mu\nu}[F_{\mu\nu},\psi]	\ ,
\end{align}
so if we consider flows for which $\lambda_{\psi}^2 = \sigma_1$ in the IR the final two terms become a total derivative. We will also fine-tune the coupling $\sigma$ such that it takes the value 1 in the IR. We can then take our IR Fermion action to be
\begin{align}
S^{(IR)}_\psi & = 	\frac{1}{2g^2}  \fint \Tr \Big(i \psi^\dag  D_t \psi  +i\chi^\dag \gamma_\mu D_\mu \psi +i\psi^\dag \gamma_\mu D_\mu \chi+ i \psi^\dag[\phi,\psi ] \Big) \ ,
\end{align}
 %which means that the final two terms of the Fermionic action are a total derivative on the constraint surface and can be dropped\footnote{This argument is a bit quick: we would generically expect the renormalisation of the two terms to be different, so the initial action would need to be finely tuned to get this result in the IR. We'll ignore this subtlety here. In particular only claim that such flows exist, not that they are generic.}. Our IR action then reduces to
 %\begin{align}
% S^{(IR)}_\psi & = 	\frac{1}{g^2} \Tr \fint  \brac{\frac{i}{2} \psi^\dag  D_t \psi  +i\chi^\dag \gamma_\mu D_\mu \psi+i\psi^\dag \gamma_\mu D_\mu \chi +i \sigma \psi^\dag[\phi,\psi ] }\ .
 %\end{align}

Combining the Bosonic and Fermionic contributions with our gauge theory, it is reasonable to propose that by deforming the original theory we flow to an infrared theory given by the action
\begin{align}
S^{(IR)} = 	\frac{1}{2g^2} \fint  \Tr&\left(  F_{t\mu} F_{t\mu} +  G_{\mu\nu}F^\pm_{\mu\nu} - D_\mu \phi^\dag D_\mu \phi \right. + i \psi^\dag  D_t \psi  \nonumber\\ &\left.\qquad +i\chi^\dag \gamma_\mu D_\mu \psi + i\psi^\dag \gamma_\mu D_\mu \chi+ i\psi^\dag [\phi,\psi] \right) \ .
\end{align}

However, this is just the theory discussed in the previous section: thus in the IR the symmetry is enhanced to the Schr\"odinger algebra.

\section{Scalar Supersymmetry}\label{sec: susy}

 The no-go discussion above can be avoided by considering a time-domain scalar supersymmetry \cite{Arav:2019tqm}. This is achieved by pairing the gauge field with a complex anticommuting $SO(4)$ vector superpartner $\psi_{\mu}$. It will prove convenient to work with off-shell supersymmetry, for which an auxiliary vector field $H_{\mu}$ is required.

 All fields take values in the adjoint of some gauge group. The action is given by 
\begin{align} \nonumber
     S_{gauge} = \inv{2g^2} \fint \Tr\Big( &F_{t\mu} F_{t\mu} - \lambda^2 D_{\nu} F_{\nu\mu} D_{\rho} F_{\rho\mu} + H_{\mu} H_{\mu}
    + 2i \psi^\dag_{\mu} D_t \psi_{\mu} \\ &- 2\lambda D_{\mu} \psi^\dag_{\nu} D_{\mu} \psi_{\nu} + 2\lambda D_{\mu}\psi^\dag_{\mu} D_{\nu}\psi_{\nu} - 4i\lambda \psi^\dag_{\mu} [F_{\mu\nu}, \psi_{\nu}] \Big) \  .
 \end{align}
A brief calculation shows this is invariant under the supersymmetry transformations
\begin{align} \label{eq:gauge susy transformations}
    \delta A_t &= 0 \nonumber \\
       \delta A_{\mu} &= i \xi^\dag \psi_{\mu} - i \psi^\dag_{\mu} \xi \nonumber\\
    \delta \psi_{\mu} &= - (F_{t\mu} + i \lambda D_{\nu}F_{\nu\mu} - i H_{\mu})\xi\nonumber  \\
    \delta H_{\mu} &= - \xi^\dag \brac{ D_t \psi_{\mu} - i \lambda D^2 \psi_{\mu} + i \lambda D_{\mu}D_{\nu} \psi_{\nu} - 2\lambda [F_{\mu\nu},\psi_{\nu}]} + h.c.\ ,
\end{align}
parameterised by a complex anticommuting scalar $\xi$. Taking the commutator of two transformations gives
 \begin{align}
    [\delta_{\xi}, \delta_{\eta}] A_{t} &= 0 \nonumber\\
    [\delta_{\xi}, \delta_{\eta}] A_{\mu} &= 2i \brac{\xi^\dag \eta - \eta^\dag \xi} F_{t\mu}  \nonumber \\
    [\delta_{\xi}, \delta_{\eta}] \psi_{\mu} &= 2i \brac{\xi^\dag \eta - \eta^\dag \xi} D_t \psi_{\mu} \nonumber\\
    [\delta_{\xi}, \delta_{\eta}] H_{\mu} &= 2i \brac{\xi^\dag \eta - \eta^\dag \xi} D_t H_{\mu}\ ,
\end{align}
 which is consistent with our expectation that the supersymmetry transformations should close to gauge-covariant translations on the fields. Our model also shares much in common with the constructions of \cite{Parisi:1982ud, Mcclain:1982fb} within the context of stochastic quantization.

\subsection{Coupling to Matter}

We'd like to couple the gauge theory described above to a Lifshitz-invariant matter theory while preserving supersymmetry. A simple way to do this is to use a complexified version of the Lifshitz Wess-Zumino model constructed in \cite{Chapman:2015wha}, which we discussed in section \eqref{sect: no-go susy} in relation to spinorial supersymmetry in two spatial dimensions. Let's look first at the free theory. The basic fields are a commuting complex scalar $X$ and two anticommuting complex scalars $(\rho,\chi)$; an auxiliary complex scalar $K$ is also included to get off-shell supersymmetry. Our free action is
\begin{align} \label{eq:phi action free} \nonumber
    S_{matter} = \fint \Big(\partial_t \bvphi \partial_t \vphi - &\lambda^2 \partial^2 \bvphi \partial^2 \vphi + i \brho \partial_t \rho - \lambda \partial_{\mu} \brho \partial_{\mu} \rho \\
    &+ i \bchi \partial_t \chi - \lambda \partial_{\mu} \bchi \partial_{\mu} \chi + K^{\dag} K\Big)\,,
\end{align}
which is invariant under the transformations
\begin{align}
\label{eq:vphi transformations}
     \delta \vphi & = \xi^\dag \rho + \chi^\dag \xi\nonumber \\
    \delta \rho & = (-i\partial_t + \lambda \partial^2)\vphi \xi + K \xi \nonumber \\
    \delta \chi & = (-i\partial_t + \lambda \partial^2)\bvphi \xi + {K}^\dag \xi \nonumber \\
    \delta K & = \xi^\dag (-i\partial_t - \lambda \partial^2) \rho + (i\partial_t - \lambda \partial^2) \chi^\dag \xi\ .
\end{align} 
The closure of the algebra on any field $\Phi$ takes the form
\begin{equation} \label{eq:closure}
    [\delta_{\xi}, \delta_{\eta}] \Phi = 2i (\xi^\dag \eta - \eta^\dag \xi) \partial_t \Phi\ ,
\end{equation}
as required for time-domain supersymmetry.

Coupling this model to the gauge theory requires us to specify how our Lie algebra $\frak{g}$ acts on the fields. Let $R$ denote a representation of $\frak{g}$, and $\Bar{R}$ the conjugate representation. These act on the vector spaces $\cal{V}$ and $\Bar{\cal{V}}$ respectively: when necessary we'll use $\alpha$ and $\balpha$ as  indices for these vector spaces. If we choose $\vphi$ to be a $\cal{V}$-valued scalar then (\ref{eq:vphi transformations}) forces us to also take $\rho$ and $K$ to be $\cal{V}$-valued, and $\chi$ to be $\Bar{\cal{V}}$-valued.

The naive gauge-covariant extension of the supersymmetry transformations (\ref{eq:vphi transformations}) obtained by the substitution
\[\partial_t \to D_t\;,\;\partial_{\mu} \to D_{\mu}\ ,\]
turns out to be inconsistent, with the algebra no longer closing on the fields. The correct extension is
 \begin{align} \label{eq:vphi transformations w gauge}
    \delta \vphi &= \xi^\dag \rho + \bchi \xi \nonumber \\
    \delta\rho &= (-i D_t + \lambda D^2)\vphi \xi + K \xi\nonumber  \\
    \delta \chi &= (-i D_t + \lambda D^2)\bvphi\xi + {K}^\dag \xi\nonumber  \\
    \delta K &= \xi^\dag\brac{ -i D_t\rho - \lambda D^2\rho - 2\lambda\psi_{\mu}^{R}D_{\mu}\vphi - \lambda (D\cdot \psi)^{R} \vphi} \\ \nonumber
    &\qquad\qquad + \brac{i D_t \chi^\dag - \lambda D^2 \chi^\dag + 2\lambda (\psi^{\dag}_{\mu})^R D_{\mu} \vphi + \lambda (D\cdot\psi^\dag)^{R}\vphi }\xi\ ,
\end{align}
which closes to
\begin{equation}
    [\delta_{\xi}, \delta_{\eta}]\Phi = 2i(\xi^\dag \eta - \eta^\dag \xi) D_t \Phi  \,,
\end{equation}
the covariant generalisation of (\ref{eq:closure}).

The natural generalisation of (\ref{eq:phi action free}) for the non-Abelian theory is
\begin{gather}
    S_{matter} = \fint \Big(D_t \bvphi_{\balpha} D_t \vphi_{\alpha} - \lambda^2 D^2 \bvphi_{\balpha} D^2 \vphi_{\alpha} + i \rho^\dag_{\balpha} D_t \rho_{\alpha} \\ \nonumber \hspace{80pt}
    + i \chi^\dag_{\alpha} D_t \chi_{\balpha} - \lambda D_{\mu} \rho^\dag_{\balpha} D_{\mu} \rho_{\alpha} - \lambda D_{\mu} \chi^\dag_{\alpha} D_{\mu} \chi_{\balpha} + {K}^\dag_{\balpha} K_{\alpha} \Big)\,.
\end{gather}
The action is not invariant under the action of (\ref{eq:gauge susy transformations}) and (\ref{eq:vphi transformations w gauge}). However, the extra terms are of the form
\begin{equation}
    \delta S_{matter} = - \delta S_{interaction}\ ,
\end{equation}
so we can restore invariance through the addition of further interaction terms. The complete supersymmetric gauge-matter action is
\begin{align}
    S &= S_{gauge} + S_{matter} + S_{interaction}\,,\nonumber \\
    S_{interaction} &= \lambda\fint \Big( (\psi^\dag_{\mu})^R_{\alpha\bbeta}\brac{D_{\mu}\bvphi_{\balpha} \rho_{\beta} - \bvphi_{\balpha} D_{\mu}\rho_{\beta} + D_{\mu}\chi_{\balpha} \vphi_{\beta} - \chi_{\balpha} D_{\mu} \vphi_{\beta}} \\ \nonumber 
    &\qquad \qquad\qquad  - (\psi_{\mu}^R)_{\alpha\bbeta} \brac{\rho^\dag_{\balpha} D_{\mu}\vphi_{\beta} - D_{\mu} \rho^\dag_{\balpha} \vphi_{\beta} - D_{\mu}\bvphi_{\balpha} \chi^\dag_{\beta} + \bvphi_{\balpha} D_{\mu} \chi^\dag_{\beta}} \\ \nonumber
    &\qquad\qquad\qquad- i (H_{\mu} - \lambda D_{\nu} F_{\nu\mu})_{\alpha\bbeta}^{R} \brac{\bvphi_{\balpha} D_{\mu} \vphi_{\beta} - D_{\mu}\bvphi_{\balpha} \vphi_{\beta} } - [(\psi^\dag_\mu)^R, \psi_{\mu}^R]_{\alpha\bbeta} \bvphi_{\balpha} \vphi_{\beta}
    \Big)\,.
\end{align}

\section{Conclusion}\label{sect: Conclusions}

In this paper we have discussed various quantum aspects of a class of five-dimensional gauge theories with a $z=2$ Lifshitz scaling. We have confirmed the computation in \cite{Horava:2008jf} that they are asymptotically free and also computed the 1-loop contributions to the $\beta$-functions arising from matter fields. In general the phase structure of these theories is quite complicated as each field can have its own $\lambda$ parameter that controls the relative strength of the spatial gradients, as noted in \cite{Iengo:2009ix}. This in turn feeds into the $\beta$-function for $g$. Depending on the values of such parameters a given theory may or may not be asymptotically free for the same matter content. 

We  also gave arguments that such theories cannot admit a spinorial supersymmetry or boosts without including Lagrange multiplier fields that restrict the dynamics. This is consistent with all the models we are aware of in more than two spatial dimensions. However, we also argued that there are RG flows starting from such UV fixed points that do lead to fixed point theories with constraints. We showed that such theories can admit a full super-Schr\"odinger symmetry that includes supersymmetry, boosts and a special conformal transformation, for example as constructed in \cite{Lambert:2018lgt,Lambert:2020jjm,Lambert:2019fne}. These are the symmetries one would expect to find from a null reduction of a six-dimensional superconformal field theory. 

We also exploited the fact that without Lorentz symmetry there is no spin-statistics theorem and hence it is consistent to consider Fermions that are scalars or vectors of the spatial rotation group. We then constructed supersymmetric Lifshitz gauge theories for which the supersymmetry is a scalar under rotations and the Fermions are vectors.  

Our interest in these non-Lorentzian theories sprung from the fact that they provide weakly coupled UV completions of five-dimensional theories with 
super-Schr\"odinger symmetry. According to various DLCQ conjectures   \cite{Aharony:1997th,Aharony:1997an} they then provide, at least at a formal level, a description of six-dimensional superconformal field theories where an additional $U(1)$ symmetry arises from the topological instanton number.
However, they also present a rich and perhaps intriguing phase structure that is  worthy of further research. Lastly, we mention that certain $\Omega$-deformed versions of these theories have been studied as they provide a way to reconstruct the  non-compact six-dimensional SCFT   \cite{Lambert:2019jwi,Lambert:2020zdc}. It is therefore of interest to explore the role of an $\Omega$ deformation in the  theories discuss here.

   \section*{Acknowledgements}

We would like to thank A. Lipstein and R. Mouland for helpful discussions. 
J.S. is supported by the STFC studentship ST/W507556/1.

\section*{Appendix A: Non-Lorentzian Dimensional Regularization}

The natural regulator for the divergent integrals arising in the 1-loop calculation is dimensional regularization, as this respects gauge-invariance. Since our theory is non-relativistic, the integrals over the energy $\omega$ and momenta $k$ need to be treated separately. If the integral over $\omega$ is convergent, then this causes no issues- we simply do this integral exactly and use standard dimensional regularisation for the remaining $k$ integrals. However, at first glance it is unclear how to extend this to the case where the $\omega$ integral is also divergent. A solution is to use the split-dimensional regularization scheme introduced in \cite{Leibbrandt:1996np}. This assigns both time and space arbitrary complex dimensions,
\begin{equation}
    dt d^4x \to d^{\sigma}t d^d x \ ,
\end{equation}
with the limits $\sigma \to 1$ and $d\to 4$ taken after all integrals are computed.

All integrals we encounter can, after Feynman parametrization, be put in the form
\begin{equation}
    \cal{J} = \int \frac{d^{\sigma} \omega d^d k}{(2\pi)^{\sigma + d}} \, \frac{(\omega^2)^a}{(\omega^2 + f(k))^b} g(k) \ .
\end{equation}
Evaluating the integral over $\omega$ gives
\begin{equation}
    \cal{J} = \frac{\Gamma\brac{b - a - \frac{\sigma}{2}} \Gamma\brac{a + \frac{\sigma}{2}}}{ (4\pi)^{\frac{\sigma}{2}} \Gamma\brac{b} \Gamma\brac{\frac{\sigma}{2}}} \int \frac{d^d k}{(2\pi)^d} \, g(k) f(k)^{a + \frac{\sigma}{2} - b} \ .
\end{equation}
As we take $\sigma\to 1$ the prefactor is finite, and the only place divergences can appear for non-zero $f(k)$ is in the integral. However, as this is already regulated by $d$ we lose nothing by setting $\sigma = 1$ before we compute the integral. This procedure allows us to ignore any divergences in the integrals over $\omega$, as they don't contribute to the renormalization of the theory at 1-loop.

\section*{Appendix B: Evaluating Traces}  
Evaluating the 1-loop contributions requires the computation of traces of the form
\begin{equation}
    X = \Tr \brac{\brac{\cal{P}_1^{-1} \Delta_1}...\brac{\cal{P}_n^{-1} \Delta_n}}\ ,
\end{equation}
where $\cal{P}_i^{-1}$ is a free field Green's function and $\Delta_i$ is a differential operator. Though this can be interpreted in terms of Feynman diagrams, it will prove simpler to directly evaluate the functional trace. In this section we will compute this in the case where all pairs $\cal{P}_i^{-1}\Delta_i$ are identical, and will therefore drop the subscript. The result for the more general case can be found using a similar approach. Translation-invariance allows us to write the Green's function as
\begin{equation}
    \cal{P}^{-1}(x,t;y,\tau) = \int \frac{d^4p d\omega}{(2\pi)^5} e^{-i(p\cdot(x-y) + \omega (t-\tau))} f(p,\omega) \ .
\end{equation}

The functional product of $n$-copies of $\cal{P}^{-1}\Delta$ is given by
\begin{align} \nonumber
    \brac{\cal{P}^{-1} \Delta}^n(x_1,x_{n+1}; \partial_x^{(n+1)},\partial_t^{(n+1)}) = \int \prod_{i=2}^n &\bigg( d^4 x_i dt_i \bigg) \prod_{j=1}^n\Big[\cal{P}^{-1}(x_j,t_j;x_{j+1},t_{j+1}) \\ & \times \Delta\brac{x_{j+1},t_{j+1};\partial_x^{(j+1)},\partial_t^{(j+1)}} \Big] \ ,
\end{align}
where we've explicitly denoted the 'dependence' on derivatives in $\Delta$. Using $\{e^{i(k\cdot x + \omega t)}\}$ as a complete basis of functions and letting $\tr$ denote traces over sets of indices (such as colour and spacetime), the trace is then
\begin{align} \nonumber
    X = \tr \int \prod_{i=1}^{n+1} \bigg(d^4x_i dt_i \bigg) &\frac{d^4 k_{n+1} d\omega_{n+1}}{(2\pi)^5} \, e^{i(k_{n+1} \cdot x_1 + \omega_{n+1} t_1)} \prod_{j=1}^n\Big[\cal{P}^{-1}(x_j,t_j;x_{j+1},t_{j+1}) \\  &\times \Delta\brac{x_{j+1},t_{j+1};\partial_x^{(j+1)},\partial_t^{(j+1)}} \Big] e^{i(k_{n+1} \cdot x_{n+1} + \omega_{n+1} t_{n+1})}\ ,
\end{align}
which after expanding $\cal{P}^{-1}$ in momentum-space becomes
\begin{equation}
\begin{split}
    X = \tr \int \prod_{i=1}^{n+1}\brac{\frac{d^4x_i dt_i d^4k_i d \omega_i}{(2\pi)^5}} e^{i(k_{n+1}\cdot x_{1} + \omega_{n+1} t_1)} \prod_{j=1}^{n} \Big[ e^{-i(k_j\cdot(x_j - x_{j+1}) + \omega_j (t_j - t_{j+1}))} \\
    \times f(k_j,\omega_j) \Delta\brac{x_{j+1},t_{j+1};\partial_s^{(j+1)},\partial_t^{(j+1)}}\Big] e^{-i(k_{n+1}\cdot x_{n+1} + \omega_{n+1} t_{n+1})} \ .
\end{split}
\end{equation}
As the derivatives in $\Delta$ act on the exponential to their right, we can make the formal substitution
\begin{equation}
    \Delta\brac{x_{j+1},t_{j+1};\partial_s^{(j+1)},\partial_t^{(j+1)}} \to \Delta\brac{x_{j+1}, t_{j+1};-ik_{j+1}, -i\omega_{j+1}}
\end{equation}
in order to rewrite the integrand solely in terms of functions of the coordinates and momenta. We observe that the only dependence on $x_1$ and $t_1$ is in the exponential factors, so we can perform the integrals over $x_1$, $t_1$, $k_1$, and $\omega_1$ to set $k_1 = k_{n+1}$ and $\omega_1 = \omega_{n+1}$. After relabelling indices, this brings $X$ into the form
\begin{equation} \hspace{-0.5cm}
    X = \tr \int \prod_{i=1}^{n}\brac{\frac{d^4x_i dt_i d^4k_i d \omega_i}{(2\pi)^5} e^{-i(k_i\cdot(x_i - x_{i+1}) + \omega_i (t_i - t_{i+1}))}} \prod_{j=1}^n \Big[ f(k_{j-1},\omega_{j-1})\, \Delta\brac{x_{j}, t_{j};-ik_{j}, -i\omega_{j}}\Big]
\end{equation}
where we take
\begin{equation}
    (k_0, \omega_0) = (k_n, \omega_n)\;,\;(x_{n+1},t_{n+1}) = (x_1, t_1) \ .
\end{equation}
Finally, expanding $\Delta$ in momentum-space as
\begin{equation}
    \Delta(x_i,t_i;-ik_i, -i\omega_i) = \int \frac{d^4 p dE}{(2\pi)^5}\, e^{-i(p_i\cdot x_i + E_i t_i )} \Tilde{\Delta}(p_i, E_i; -ik_i, -i\omega_i)
\end{equation}
allows us to perform the coordinate integrals; our final expression for $X$ is then
\begin{equation} \hspace{-1.2cm}
    X = \tr \int \prod_{i=1}^n \brac{ \frac{d^4 k_i d\omega_i\, d^4p_i dE_i}{(2\pi)^5} \, \delta^{(4)}(k_i - k_{i-1} + p_i) \delta(\omega_i - \omega_{i-1} + E_i) f(k_{i-1}, \omega_{i-1}) \Tilde{\Delta}(p_i,E_i; -i k_i, -i \omega_i)} \ .
\end{equation}

\end{document}